\documentclass{article}
\usepackage[table]{xcolor}
\usepackage{spconf,graphicx}
\usepackage[hyphens]{url}
\usepackage[]{hyperref}
\usepackage{tikz,tikz-qtree}
\usepackage{booktabs}
\usepackage{multirow}

\hypersetup{breaklinks=true}
\title{The Second DiCOVA Challenge: Dataset and performance analysis for Diagnosis of COVID-19 using acoustics}
\name{%
    \parbox{\linewidth}{\centering
      Neeraj Kumar Sharma,
      Srikanth Raj Chetupalli,
      Debarpan Bhattacharya,\\
      Debottam Dutta,
      Pravin Mote,
      Sriram Ganapathy
    }%
}
\address{LEAP lab, Indian Institute of Science, Bangalore-560012, India. \\
Email: sriramg@iisc.ac.in}

\begin{document}
\ninept

\maketitle

\begin{abstract}
The Second Diagnosis of COVID-19 using Acoustics (DiCOVA) Challenge aimed at accelerating the research in acoustics based detection of COVID-19, a topic at the intersection of acoustics,  signal processing, machine learning, and healthcare. This paper presents the details of the challenge, which was an open call for researchers to analyze a dataset of audio recordings consisting of breathing, cough and speech signals. This data was collected from individuals with and without COVID-19 infection, and the task in the challenge was a two-class classification. The development set audio recordings were collected from $965$ ($172$ COVID-19 positive) individuals, while the evaluation set contained data from $471$ individuals ($71$ COVID-19 positive). The challenge featured four tracks, one associated with each sound category of cough, speech and breathing, and a fourth fusion track. A baseline system  was also released to benchmark the participants. In this paper, we present an overview of the challenge, the rationale for the data collection and the baseline system. Further, a performance analysis for the systems submitted by the $16$  participating teams in the leaderboard is also presented. 
\end{abstract}
\section{Introduction}
\label{sec:intro}
Owing to the cost of machinery, expertise, and time, the molecular testing approaches, namely, the reverse transcription polymerase chain reaction (RT-PCR) test and the rapid antigen test (RAT) cannot be easily scaled up and deployed in a short time \cite{mercer2021testing}. This results in a bottleneck in containing the spread of COVID-19, and has led to research on alternate testing methodologies \cite{kevadiya2021diagnostics}. Some of these include nanomaterial based bio-sensing of SARS-CoV-2 virus \cite{chacon2020optimized} and radiographic imaging based on computed tomography (CT) \cite{hosseiny2020radiology}, X-ray (CTX) \cite{ctx_study}, and ultrasound \cite{poggiali2020can} to categorize the health status of lungs.

The primary symptoms of COVID-19 disease are fever, sore throat, cough, chest, muscle pain, and dyspnoea. Further, the pathogenesis of COVID-19 indicates minor to acute infection in the respiratory system during the onset of the disease. This has garnered interest in the speech and audio research community, and there have been several studies \cite{imran2020ai4covid, orlandic2020coughvid, brown2020exploring, laguarta} gathering insights on the possibility of acoustics based COVID-19 diagnosis. Such an approach can provide a point-of-care, rapid, easy to use, and cost-effective tool to help contain COVID-19 spread. This paper discusses the second in the series of DiCOVA open challenges, an attempt to benchmark the acoustic based diagnosis research among various groups.
\begin{figure*}[t]
    \centering
    \input{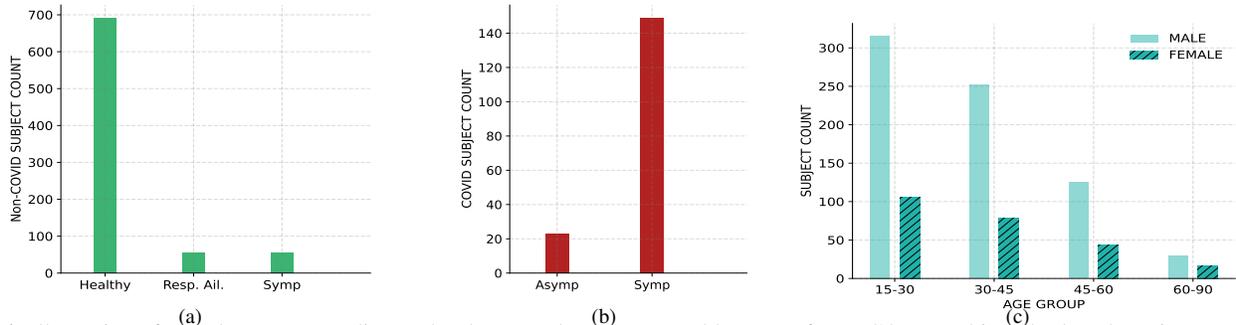}
    \vspace{-0.25in}
    \caption{Illustration of metadata corresponding to development dataset. (a) Health status of Non-COVID subjects broken down into categories of healthy (no symptoms), pre-existing respiratory ailment (asthma, chronic lung disease, pneumonia), and symptoms (cold, cough, fever, loss of taste or smell); (b) COVID-19 status of the subjects; (c) Pooled subject gender and age distribution.}
    \label{fig:dev_set_metadata}
\vspace{-0.25in}
\end{figure*}

The DiCOVA challenge series is designed with a curated development dataset taken from the crowd-sourced Coswara dataset \cite{sharma2020coswara}. The challenge dataset, with labels and a baseline system, are released and researchers are invited to develop machine learning systems that perform well on a blind test set. A leader-board style ranking is created for the evaluation on blind test set. The first DiCOVA Challenge\footnote{\url{http://dicova2021.github.io/}} was launched earlier~\cite{dicova}, which had garnered participation from $28$ teams, coming from both academia and industry. Among these teams, $21$ teams surpassed the baseline system performance. A summary of the challenge  is provided in \cite{sharma2021towards}. 

Inspired by the interest of the research community, and the increasing demand for fast, remote and accurate COVID-19 testing methodologies, we had launched the second DiCOVA challenge\footnote{\url{http://dicovachallenge.github.io/}} on 12-Aug-2021. The key considerations over the previous challenge that  motivated us to conduct the second DiCOVA challenge are the following,  

\begin{itemize}
    \item Since the closing of the first DiCOVA Challenge, there was a spike in daily global COVID-19 cases (Apr-May, 2021). The spike has been attributed to the new strains of the virus. This has enabled us to increase the data set size for the Second DiCOVA Challenge.
    \item In addition to the cough sound category, the challenge brings to focus two additional sound categories, namely, breathing and speech. A leaderboard-style evaluation on blind test set is built for four tracks, one  associated  with  each sound category (that is, breathing, cough, and speech) and a fourth fusion track allowing experimentation with combinations of the individual sound categories.
    \item Recently, multiple open datasets have been released to the public by different research groups. These include COVID-19 Sounds dataset \cite{covid19sounddetector} by University of Cambridge (UK), Buenos Aires COVID-19 Cough dataset \cite{badata} by Cabinet Ministers (Argentina), COUGHVID dataset \cite{orlandic2020coughvid} by EPFL University (Switzerland), and COVID-19 Open COUGH dataset \cite{virufy_set} by Virufy (US).
The participants were encouraged to use these datasets for enhancing model training and analysis.
\end{itemize} 

In this paper, we present an overview of the Second DiCOVA Challenge which spanned $51~$days and concluded on 01-Oct-2021.

\section{Literature Review}
\label{sec:lit_review}
In a study by Imran et al.~\cite{imran2020ai4covid}, a four class classifier was designed to detect healthy, pertussis, bronchitis, and COVID-19 individuals. On a privately collected cough sound dataset from hospitals, they report a sensitivity of $~94\%$ (and ~$91\%$ specificity) using a convolutional neural network (CNN) architecture with mel-spectrogram feature input. Other studies have focused on the binary task of COVID-19 detection only. Agbley et al. \cite{agbley2020wavelet} reported $81\%$ specificity (at $43\%$ sensitivity) on a subset of the COUGHVID dataset \cite{orlandic2020coughvid}. Laguarte et al. \cite{laguarta} used  a privately collected dataset of COVID-19 infected individuals and report an area under the-receiver operating characteristic curve (AUC-ROC) performance of $97.0\%$. Andreu-Perez et al. \cite{9361107} created a controlled dataset by collecting cough sound samples from patients visiting hospitals, and report $98.8\%$ AUC-ROC. A few studies have explored using the breathing and voice sounds as well. Brown et al.~\cite{brown2020exploring} created a dataset through crowd-sourcing, and analyzed COVID-19 detection. The authors report a performance between $80-82\%$ AUC-ROC. Han et al~\cite{han_voice_symptoms} proposed using voice samples and demonstrate $77\%$ AUC-ROC. Further, the use of symptoms along with voice provides a $2\%$ improvement in the AUC-ROC.

While these studies are encouraging, there are several limitations \cite{coppock2021_grains_covid}. Particularly, $(i)$ different COVID-19 patient population (with varied sizes) is used in each study, $(ii)$ use of different performance evaluation methodologies across the studies, and $(iii)$ lack of common data and reproducibility. The Second DiCOVA Challenge was aimed at encouraging research groups to design and evaluate their classification system on a common dataset, using the same performance metrics. This addresses $(i)$ and $(ii)$, and helps us to benchmark the designed systems against a baseline system. To address $(iii)$ we encouraged the participants to submit detailed system reports and open source the developed systems.

\section{Dataset}
\label{sec:dataset}
\noindent The challenge dataset is derived from the Coswara dataset \cite{sharma2020coswara}, a crowd-sourced dataset of sound recordings\footnote{\url{https://coswara.iisc.ac.in/}}. The volunteers from across the globe, age groups and health conditions were encouraged to record their sound data in a quiet environment using a web connected device (like smartphone or computer). Through the website, the subjects first provide demographic information like age and gender. This is followed by a short questionnaire to record their health status, including symptoms, pre-existing respiratory ailments, and co-morbidity, if any. Subsequently, their COVID-19 status is recorded by asking if they are currently COVID-19 positive, recovered, exposed to COVID-19 patients through primary contacts, or healthy. After collecting this information as metadata, the subjects record their acoustic data corresponding to $9$ audio categories, namely, $(a)$ shallow and deep breathing ($2~$ types), $(b)$ shallow and heavy cough ($2$ types), $(c)$ sustained phonation of vowels [\ae] (as in bat), [i] (as in beet), and [u] (as in boot) ($3$ types), and $(d)$ fast and normal pace number counting ($2$ types). The whole process takes $5-7$~minutes. The dataset collection protocol was approved by the Human Ethics Committee of the Indian Institute of Science, Bangalore (India).

The Second DiCOVA Challenge used a subset of the Coswara dataset, sampled from the data collected between Apr-$2020$ and Jul-$2021$. The sampling included only the age group of $15-90$ years. The subjects with health status of ``recovered'' (that is, already recovered from COVID infection during the time of recording) and ``exposed'' (suspecting exposure to the virus through primary contacts) were not included in the dataset.  Further, subjects with audio recording duration less than $500$~msec were discarded. Only three sound categories are considered in the challenge. These correspond to breathing-deep, cough-heavy, and counting-normal, and for brevity are referred to as breathing, cough, and speech, respectively. The resulting curated subject pool was divided into the following two groups. 

\begin{itemize}
\item \textbf{Non-COVID}: Subjects self-declared as healthy or having COVID-19 like symptoms (such as cold, cough, fever, muscle pain or fatigue, loss of taste or smell) or pre-existing respiratory ailments (such as asthma, pneumonia, chronic lung disease) but were not tested positive for COVID-19.
\item \textbf{COVID}: Subjects self-declared as COVID-19 positive (asymptomatic or symptomatic with mild/moderate infection)
\end{itemize} 
\subsection{Development data}
The development dataset release is composed of audio records from $965$~($172$~COVID) subjects. This results in $965$~(subjects)$\times3$ (sound categories) audio recordings. An illustration of the metadata of the subject pool is provided in Figure~\ref{fig:dev_set_metadata}.
About $70$\% of the subjects were male. The majority of the participants lie in the age group of $15-45$ years. In the non-COVID subject pool, close to $86\%$ are healthy without respiratory ailments or COVID-19-like symptoms. In the COVID subject pool, close to $87\%$ are symptomatic. In the development set, $17.2\%$ subjects belong to COVID class. This represents an imbalanced dataset, reflecting the typical real-world scenario in the design of point-of-care-tests (POCTs) for COVID-19. The development dataset release also contains a five fold ($80-20\%$) train-validation split of this dataset. This provides an option to the participants for exploring hyper-parameter tuning in their models.
\subsection{Evaluation data}
For evaluation, a blind test set consisting of $471$ audio files $\times 3$ sound categories was provided. This contained $71$ COVID-19 positive individuals while the remaining subjects were Non-COVID. The age, and gender distribution on the blind test set was matched with those in the development set. The participating teams were required to upload their probability scores for the blind test set audio files to an online evaluation website portal\footnote{\url{https://competitions.codalab.org/competitions/34801}}. This portal was equipped to compute the performance using the evaluation metrics (defined in Section~\ref{sec:metrics}) and rank order the teams performance on a common leader-board. This was done  real-time. Every team received a maximum of $15$ tickets-per-track for submitting scores to the leader-board. Following the evaluation, the teams have been provided with the labels and the metadata corresponding to the blind test set. 

\subsection{Audio Specifications}
\noindent All audio recordings were re-sampled to $44.1$~kHz and compressed as FLAC (Free Lossless Audio Codec) format for the ease of distribution. The duration of audio in each track corresponded to $4.62~$hrs for Track-1, $1.68~$hrs for Track-2, and $3.93~$hrs for Track-3.

\section{Challenge Tasks}
\label{sec:tasks}
The challenge required designing a binary classifier to detect COVID/Non-COVID health status of subjects using sound categories corresponding to each track. Every registered team was  provided with the development dataset to facilitate training and design of classification models. The teams were free to use any dataset except the publicly available Coswara dataset\footnote{\url{https://github.com/iiscleap/Coswara-Data}} for data augmentation purposes. 

Alongside the development dataset and blind test set, a baseline software implementation was also shared with the participants. This provided a data analysis code pipeline and model training recipe on the development dataset with five-fold validation. Post challenge, that is, $01$-Oct-$2021$, teams submitted their system report describing the designed models to the challenge organizers.
Every team signed a terms and condition document stating the data licenses, the fair use of data and  the rules of the challenge.
\begin{figure*}[!t]
    \centering
    \input{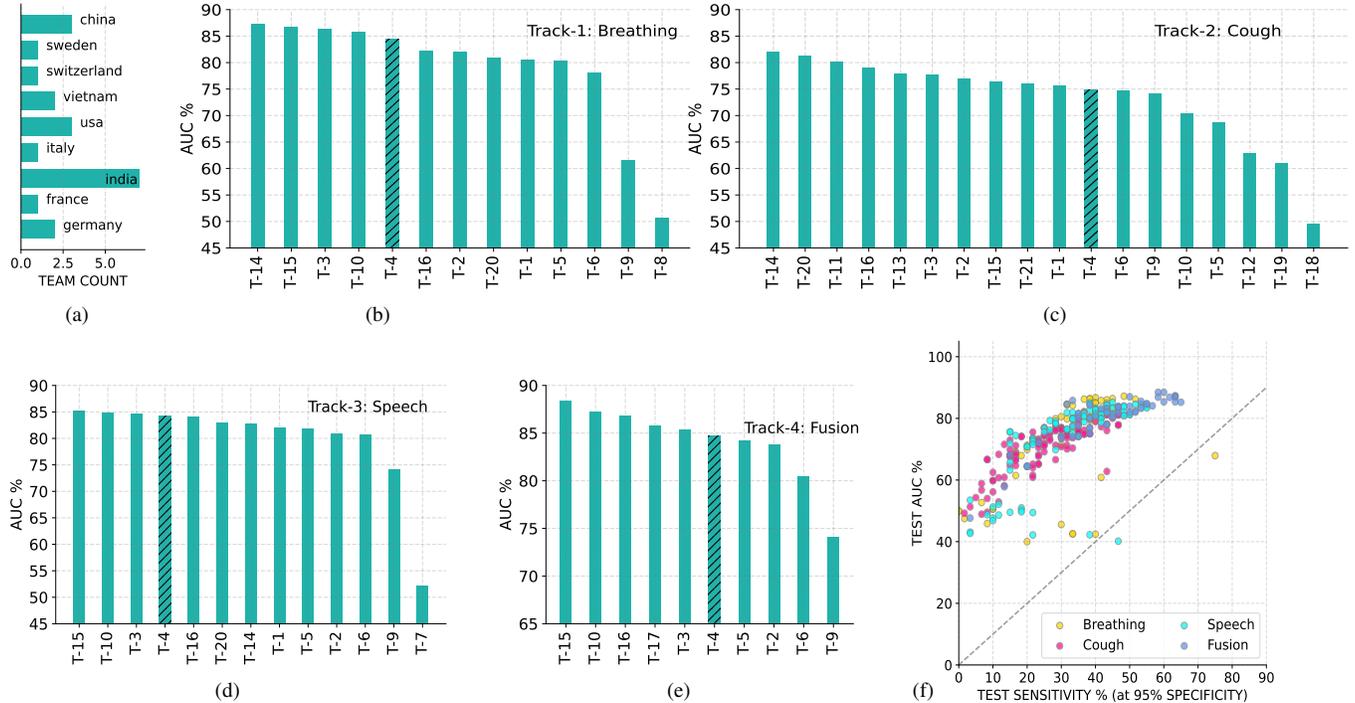}
    \vspace{-0.25in}
    \caption{Illustration of results of different teams, indicated by T-$n$ where $n$ denotes the index after sorting teams by their names. (a) Distribution of teams based on country of origin. (b-e) Best Blind Test AUC (under the ROC) \% posted by different teams on the leaderboard (ordered in descending order). (f) Scatter plot of Blind Test AUC versus Sensitivity (at $95\%$ Specificity) for all submissions made to the evaluation portal. Note: Team T-4 (hatched bar) stands for the baseline.}
    \label{fig:leaderboard}
\vspace{-0.25in}
\end{figure*}
\subsection{Evaluation Metrics}\label{sec:metrics}
\noindent As the dataset was imbalanced, we choose not to use accuracy as an evaluation metric. Each team submitted the COVID probability score, with a higher value indicating a higher likelihood of COVID infection, for the list of validation and test audio recordings. The web interface used the scores and the ground truth labels to compute the receiver operating characteristics (ROC) curve. The curve is obtained by varying the decision threshold with a step size of $0.0001$ and obtaining the specificity (true negative rate) and sensitivity (true positive rate) at every threshold value. The area under the resulting ROC curve, AUC-ROC, was used as the primary performance metric. The area was computed using the trapezoidal method. Further, the sensitivity at a specificity of $95\%$ was used as a secondary evaluation metric. For brevity, we will refer to AUC-ROC as AUC in the rest of the paper.
\begin{table}[]
\centering
\rowcolors{2}{white}{gray!10}
\begin{tabular}{@{}llcccc@{}}
\toprule
\multirow{2}{*}{\textbf{Validation}} &  & \multicolumn{4}{c}{\textbf{AUC-ROC Performance (in \%)}} \\ \cmidrule(l){3-6} 
 &  & \textbf{Breathing} & \textbf{Cough} & \textbf{Speech} & \textbf{Fusion} \\ \midrule
fold-0 &  & 74.8 & 71.8 & 75.4 & 77.3 \\
fold-1 &  & 73.9 & 78.2 & 87.2 & 82.4 \\
fold-2 &  & 74.3 & 77.2 & 80.6 & 81.8 \\
fold-3 &  & 80.0 & 74.0 & 78.2 & 80.3 \\
fold-4 &  & 83.2 & 74.9 & 79.5 & 86.6 \\ \midrule
\textbf{Avg. Validation} &  & \textbf{77.3} & \textbf{75.2} & \textbf{80.2} & \textbf{81.7} \\ \midrule
\textbf{Test} &  & \textbf{84.5} & \textbf{74.9} & \textbf{84.3} & \textbf{84.7} \\\bottomrule
\end{tabular}
\caption{Baseline system performance on the validation folds in the development dataset, and the blind test dataset.}
\label{table:perfm_val}
    \vspace{-0.25in}
\end{table}

\section{Baseline System}
\label{sec:baseline}
The different processing stages in the baseline system are described below. The same baseline system setup was used for all tracks.
\\
\textit{\bf{Pre-processing:}} The audio sample was normalized to lie between $\pm~1$. This was followed by discarding low activity regions from the signal. Using a sound sample activity detection threshold of $0.01$, and a buffer size of $50~$msec on either side of a sample, any audio region with sample values greater than the threshold was retained.
\\
\textit{\bf{Feature extraction:}} The log mel-spectrogram features were extracted using short-time windowed segments of size $1024$ samples ($23.2~$msec) and temporal hop of $441$ samples ($10$~ms), and a $64$ mel filters. This resulted in a $64\times N_k$ dimensional feature matrix for the $k^{th}$ sound file, where $N_k$ represents the number of short-time frames.
The mel-spectrogram features were appended with the first and second order temporal derivatives. The resulting $192\times N_k$ dimensional features were file-level mean-variance normalized.
\\
\textit{\bf{Classifier:}} The initial experimentation with the  traditional classifiers such as logistic regression and random forest gave a performance in the range of $60-70\%$ AUC on the validation folds for all the tracks. With the aim to provide a competent baseline, we opted using a deep learning framework with a cascade of two bi-directional long-short term memory (BiLSTM) and a fully connected layer. The initial BiLSTM layers with $128$ units were used to model the long-term dependencies in the audio signal. The output of the BiLSTM layers are of dimension $256\times T$. This is fed to a pooling layer which performs averaging along the time dimension to generate a sequence level embedding of $256\times1$. This output is fed to  a fully connected feedforward layer of $64$ nodes and a $\tanh(\cdot)$ non-linearity. The final layer resembles a two class logistic regression model with $64$ input nodes.
\\
\noindent \textit{\bf{Training:}} For training the classifier, contiguous segments were extracted, with a $10$ frame stride, from the features matrix to obtain $192\times T$ fixed dimensional feature  representations. We choose $T$ as $51$ in the baseline system. The label of each chunk is the same as that of the audio file. Each mini-batch is composed of $1024$ feature matrices of size $192\times T$, randomly sampled from different audio files such that the proportion of COVID and Non-COVID labels is balanced. This oversampling of the  minority class is done to overcome the limitation of class imbalance. The binary cross entropy (BCE) loss, Adam optimizer with an initial learning rate of $0.0001,$ and $\ell_2$ regularization set to $0.0001$, were used to train the classifier. The learning rate was reduced by a factor of $10$ for a patience parameter set to three epochs. A dropout factor of $0.1$ was applied to the outputs of the first BiLSTM layer and the feedforward layer.
\\
\textit{\bf{Inference:}} Given an audio recording, $192\times T$ mel-spectrogram feature matrices (with a stride of $10$ frames) were extracted (similar to the training stage). These were input to the trained classifier and the output probability scores were obtained for each chunk. The average of the  probability scores from each segment was output as the COVID probability score of the audio file. On the blind test set, for each sound category, we draw inference by averaging the score obtained using a model trained on each validation fold.
\\
\textit{\bf{Fusion:}} Three classifiers are trained separately for the three trackwise sound categories. A final prediction at the subject level is obtained as the arithmetic mean of the COVID probability scores for the different audio sound categories. 

\section{Results}
For the baseline system, the fold-wise AUCs for the sound categories are shown in Table~\ref{table:perfm_val}. The AUC performance is better than chance ($50\%$ AUC) for all the sound categories.  An average AUC performance of $77.3\%,~75.2\%$, and $80.2\%$ was obtained for breathing, cough, and speech categories, respectively. The best average AUC of $81.67\% $was obtained with the fusion of the categories.  Amongst the three sound categories, the best test set AUC of $84.5\%$, was obtained for the breathing sound category. This was followed by speech and cough. 

Sixteen teams actively competed on the challenge leader-board. These came from different countries (see Figure~\ref{fig:leaderboard}(a)). An illustration of leader-board rankings for each sound category is shown in Figure~\ref{fig:leaderboard}(b-e). A total of $10$ teams outperformed the baseline performance in Track-2 (Cough). This number was lower for other tracks ($4$ in Track-1 (Breathing), $3$ in Track-3 (Speech), and $5$ in Track-4, respectively). A best AUC performance of $87.2\%$, $82.0\%$, $85.2\%$, and $88.4\%$ was reported in the breathing, cough, speech, and fusion tracks, respectively. Figure~\ref{fig:leaderboard}(f) depicts a scatter plot of AUC\% versus sensitivity (at $95\%$ specificity) obtained using all the $320$ submissions made to the evaluation portal.

The team T-15, top performer in Track-3 and Track-4, employed the BiLSTM architecture similar to the baseline system but with a novel initialization obtained by averaging model parameters across sound categories. They also used high-level features (that is, wav2vec2.0 \cite{baevski2020wav2vec}) alongside the MFCCs. The team T-14, top performer in Track-1 and Track-2, employed classification models based on random forests and multi-layer perceptron, and acoustic features such as RelAtive SpecTrAl-Perceptual Linear Prediction (RASTA-PLP) \cite{hermansky1994rasta}. The system reports of teams agreeing to make them public is provided in the challenge website.

Post challenge, we did an additional experiment. For each track, we averaged the normalized probability scores submitted by the top three teams for the blind test set, and considered this as a hypothetical "pooled system". This system showed slight improvement over all teams. The obtained AUCs were $88.8\%$, $84.5\%$, $86.7$, and $90.4\%$ for Track-1, 2, 3, and 4, respectively.   

\vspace{-0.15in}
\section{Conclusion}
\label{sec:conclusion}
The Second DiCOVA Challenge introduced multiple categories of acoustic signals like cough, breathing and speech  for developing a COVID-19 diagnosis approach. A curated dataset, a baseline system, and a blind test set were provided to all participants. Out of a total of $63$ registered teams, $16$ competed actively in the final leader-board. Several teams surpassed the strong baseline performance. The results strengthen the hypothesis on presence of acoustic signature of COVID-19 in respiratory sound signals, and encourage development of acoustic based point-of-care testing tools.
\vspace{-0.15in}
\section{Acknowledgement}
\noindent The authors would like to express gratitude to Ananya Muguli, Prashant Krishnan, and Rohit Kumar for help in challenge logistics, and Anand Mohan, Lance Pinto, Chandra Kiran, and Murali Alagesan for coordination in audio data collection.

\bibliographystyle{IEEEbib}
\bibliography{mybib}

\begin{thebibliography}{10}

\bibitem{mercer2021testing}
Tim~R Mercer and Marc Salit,
\newblock ``Testing at scale during the covid-19 pandemic,''
\newblock {\em Nature Reviews Genetics}, vol. 22, no. 7, pp. 415--426, 2021.

\bibitem{kevadiya2021diagnostics}
Bhavesh~D Kevadiya, Jatin Machhi, Jonathan Herskovitz, Maxim~D Oleynikov,
  Wilson~R Blomberg, Neha Bajwa, Dhruvkumar Soni, Srijanee Das, Mahmudul Hasan,
  Milankumar Patel, et~al.,
\newblock ``Diagnostics for {SARS-CoV-2} infections,''
\newblock {\em Nature materials}, pp. 1--13, 2021.

\bibitem{chacon2020optimized}
Julio~C Chac{\'o}n-Torres, C~Reinoso, Daniela~G Navas-Le{\'o}n, Sarah
  Brice{\~n}o, and Gema Gonz{\'a}lez,
\newblock ``Optimized and scalable synthesis of magnetic nanoparticles for
  {RNA} extraction in response to developing countries' needs in the detection
  and control of {SARS-CoV-2},''
\newblock {\em Scientific reports}, vol. 10, no. 1, pp. 1--10, 2020.

\bibitem{hosseiny2020radiology}
Melina Hosseiny, Soheil Kooraki, Ali Gholamrezanezhad, Sravanthi Reddy, and Lee
  Myers,
\newblock ``Radiology perspective of coronavirus disease 2019 (covid-19):
  lessons from severe acute respiratory syndrome and middle east respiratory
  syndrome,''
\newblock {\em American Journal of Roentgenology}, vol. 214, no. 5, pp.
  1078--1082, 2020.

\bibitem{ctx_study}
Stephanie Stephanie, Thomas Shum, Heather Cleveland, Suryanarayana~R. Challa,
  Allison Herring, Francine~L. Jacobson, Hiroto Hatabu, Suzanne~C. Byrne, Kumar
  Shashi, Tetsuro Araki, Jose~A. Hernandez, Charles~S. White, Rydhwana Hossain,
  Andetta~R. Hunsaker, and Mark~M. Hammer,
\newblock ``Determinants of chest radiography sensitivity for covid-19: A
  multi-institutional study in the united states,''
\newblock {\em Radiology: Cardiothoracic Imaging}, vol. 2, no. 5, pp. e200337,
  2020,
\newblock PMID: 33778628.

\bibitem{poggiali2020can}
Erika Poggiali, Alessandro Dacrema, Davide Bastoni, Valentina Tinelli, Elena
  Demichele, Pau Mateo~Ramos, Teodoro Marcian{\`o}, Matteo Silva, Andrea
  Vercelli, and Andrea Magnacavallo,
\newblock ``Can lung us help critical care clinicians in the early diagnosis of
  novel coronavirus (covid-19) pneumonia?,''
\newblock {\em Radiology}, vol. 295, no. 3, pp. E6--E6, 2020.

\bibitem{imran2020ai4covid}
Ali Imran, Iryna Posokhova, Haneya~N. Qureshi, Usama Masood, Muhammad~Sajid
  Riaz, Kamran Ali, Charles~N. John, MD~Iftikhar Hussain, and Muhammad Nabeel,
\newblock ``{AI4COVID-19: AI enabled preliminary diagnosis for COVID-19 from
  cough samples via an app},''
\newblock {\em Informatics in Medicine Unlocked}, vol. 20, pp. 100378, 2020.

\bibitem{orlandic2020coughvid}
Lara Orlandic, Tomas Teijeiro, and David Atienza,
\newblock ``{The COUGHVID crowdsourcing dataset, a corpus for the study of
  large-scale cough analysis algorithms},''
\newblock {\em Scientific Data}, vol. 8, no. 1, pp. 1--10, 2021.

\bibitem{brown2020exploring}
Chlo\"{e} Brown, Jagmohan Chauhan, Andreas Grammenos, Jing Han, Apinan
  Hasthanasombat, Dimitris Spathis, Tong Xia, Pietro Cicuta, and Cecilia
  Mascolo,
\newblock ``Exploring automatic diagnosis of {COVID-19} from crowdsourced
  respiratory sound data,''
\newblock in {\em Proc.~26th ACM SIGKDD Intl. Conf. Knowledge Discovery \& Data
  Mining}, New York, NY, USA, 2020, p. 3474–3484, Association for Computing
  Machinery.

\bibitem{laguarta}
Jordi Laguarta, Ferran Hueto, and Brian Subirana,
\newblock ``{{COVID}-19} artificial intelligence diagnosis using only cough
  recordings,''
\newblock {\em IEEE Open Journal of Engineering in Medicine and Biology}, vol.
  1, pp. 275--281, 2020.

\bibitem{sharma2020coswara}
Neeraj Sharma, Prashant Krishnan, Rohit Kumar, Shreyas Ramoji, Srikanth~Raj
  Chetupalli, R~Nirmala, Prasanta~Kumar Ghosh, and Sriram Ganapathy,
\newblock ``Coswara -- a database of breathing, cough, and voice sounds for
  {{COVID}-19} diagnosis,''
\newblock in {\em Proc. Interspeech}, 2020, pp. 4811--4815.

\bibitem{dicova}
Ananya Muguli, Lancelot Pinto, R~Nirmala, Neeraj Sharma, Prashant Krishnan,
  Prasanta~Kumar Ghosh, Rohit Kumar, Shrirama Bhat, Srikanth~Raj Chetupalli,
  Sriram Ganapathy, Shreyas Ramoji, and Viral Nanda,
\newblock ``{DiCOVA Challenge: Dataset, Task, and Baseline System for COVID-19
  Diagnosis Using Acoustics},''
\newblock in {\em Proc. Interspeech 2021}, 2021, pp. 901--905.

\bibitem{sharma2021towards}
Neeraj~Kumar Sharma, Ananya Muguli, Prashant Krishnan, Rohit Kumar,
  Srikanth~Raj Chetupalli, and Sriram Ganapathy,
\newblock ``Towards sound based testing of covid-19--summary of the first
  diagnostics of covid-19 using acoustics (dicova) challenge,''
\newblock {\em arXiv preprint arXiv:2106.10997}, 2021.

\bibitem{covid19sounddetector}
``{Cambridge University, UK - COVID-19 Sounds App},''
  \url{https://www.covid-19-sounds.org/en/blog/data_sharing.html}, 2020,
\newblock [Online; accessed 16-Aug-2021].

\bibitem{badata}
``{Buenos Aires COVID-19 Cough Data Dataset},''
  \url{https://data.buenosaires.gob.ar/dataset/tos-covid-19/}, 2021,
\newblock [Online; accessed 16-Aug-2021].

\bibitem{virufy_set}
``{Virufy COVID-19 Open Cough Dataset},''
  \url{https://github.com/virufy/virufy-data}, 2021,
\newblock [Online; accessed 04-Jun-2021].

\bibitem{agbley2020wavelet}
Bless Lord~Y Agbley, Jianping Li, Aminul Haq, Bernard Cobbinah, Delanyo
  Kulevome, Priscilla~A Agbefu, and Bright Eleeza,
\newblock ``Wavelet-based cough signal decomposition for multimodal
  classification,''
\newblock in {\em 17th Intl. Computer Conference on Wavelet Active Media
  Technology and Information Processing)}. IEEE, 2020, pp. 5--9.

\bibitem{9361107}
Javier Andreu-Perez, Humberto Perez-Espinosa, Eva Timonet, Mehrin Kiani,
  Manuel~Ivan Giron-Perez, Alma~B. Benitez-Trinidad, Delaram Jarchi, Alejandro
  Rosales, Nick Gkatzoulis, Orion~F. Reyes-Galaviz, Alejandro Torres, Carlos
  Alberto Reyes-Garcia, Zulfiqar Ali, and Francisco Rivas,
\newblock ``A generic deep learning based cough analysis system from clinically
  validated samples for point-of-need {{COVID}-19} test and severity levels,''
\newblock {\em IEEE Trans. Services Computing}, pp. 1--1, 2021.

\bibitem{han_voice_symptoms}
Jing Han, Chloë Brown, Jagmohan Chauhan, Andreas Grammenos, Apinan
  Hasthanasombat, Dimitris Spathis, Tong Xia, Pietro Cicuta, and Cecilia
  Mascolo,
\newblock ``Exploring automatic covid-19 diagnosis via voice and symptoms from
  crowdsourced data,''
\newblock in {\em IEEE Intl. Conf. Acoustics, Speech and Signal Processing
  (ICASSP)}, 2021, pp. 8328--8332.

\bibitem{coppock2021_grains_covid}
Harry Coppock, Lyn Jones, Ivan Kiskin, and Bj{\"o}rn Schuller,
\newblock ``Covid-19 detection from audio: seven grains of salt,''
\newblock {\em The Lancet Digital Health}, vol. 3, no. 9, pp. e537--e538, 2021.

\bibitem{baevski2020wav2vec}
Alexei Baevski, Henry Zhou, Abdelrahman Mohamed, and Michael Auli,
\newblock ``wav2vec 2.0: A framework for self-supervised learning of speech
  representations,''
\newblock {\em arXiv preprint arXiv:2006.11477}, 2020.

\bibitem{hermansky1994rasta}
Hynek Hermansky and Nelson Morgan,
\newblock ``{RASTA} processing of speech,''
\newblock {\em IEEE Trans. Speech and Audio Proc.}, vol. 2, no. 4, pp.
  578--589, 1994.

\end{thebibliography}

\end{document}